\documentstyle[12pt]{article}
\begin{document}

\newcommand{\be}{\begin{equation}}
\newcommand{\ee}{\end{equation}}
\newcommand{\ba}{\begin{eqnarray}}
\newcommand{\ea}{\end{eqnarray}}
\newcommand{\pa}{{\partial}}
\newcommand{\e}{{\varepsilon}}
\newcommand{\de}{{\delta}}
\newcommand{\thl}{\theta_L}
\newcommand{\thr}{\theta_R}
\newcommand{\th}{{\theta}}
\newcommand{\dep}{{\delta_\varepsilon}}
\newcommand{\deta}{{\delta_\eta}}
\newcommand{\crr}{\\ & & \nonumber \\}
\newcommand{\pht}{{\tilde\phi}}

\newcommand{\jm}{J_-}
\newcommand{\jp}{J_+}

\begin{center}
\hfill ITP-SB-94-22\\
\hfill hep-th/9406063\\
\vskip .75in

{\large COMPLEX STRUCTURES, DUALITY AND WZW-MODELS IN EXTENDED
SUPERSPACE}\\
\vskip .5in
Ivan T. Ivanov\footnotemark
\footnotetext{e-mail address: iti@insti.physics.sunysb.edu},
Byungbae Kim\footnotemark
\footnotetext{e-mail address: bkim@insti.physics.sunysb.edu},
and Martin Ro\v{c}ek\footnotemark \\
\footnotetext{e-mail address: rocek@insti.physics.sunysb.edu}
Institute for Theoretical Physics,  \\
State University of New York at Stony Brook, \\
Stony Brook, NY 11794-3840  USA \\
\vspace{2cm}
\end{center}
\begin{abstract}
We find the complex structure on the dual of a complex target space.
For $N=(2,2)$ systems, we prove that the space orthogonal to the
kernel of the commutator of the left and right complex structures is
{\em always} integrable, and hence the kernel is parametrized by chiral
and twisted chiral superfield coordinates. We then analyze the
particular case of $SU(2)\times SU(2)$, and are led to a new $N=2$
superspace formulation of the $SU(2)\times U(1)$ WZW-model.
\end{abstract}
\vfill
\newpage

\section{Introduction}
The low energy sector of a string theory is conveniently described
by studying maps from a Riemann surface to a target space.
Classically, this gives rise to generalized harmonic map problems,
and at the quantum level, this gives a description of $d=2$ quantum
field theory.  Duality \cite{FT,TB}\footnote{See also \cite{HKLR}
for a simple description of some aspects, and \cite{RV} for a
discussion of global issues.} is a map between different target space
descriptions of the same field theory, and has given a great deal of
insight into the nonperturbative structure of the space of string
backgrounds \cite{duality,more,GR}. Here we study how duality
acts on complex structures on the target space.  Specifically,
given a target space with an hermitian complex structure preserved
by the appropriate connection \cite{GHR}, we find the dual complex
structure that is similarly compatible with respect to the
metric and connection on the dual target space.

On target spaces with torsion in
$N=2$ supersymmetric theories, there are two relevant complex
structures \cite{GHR} (see also \cite{BLR,R}).  When they
commute, an $N=2$ superspace description of the model can be given in
terms of a potential that is a function of {\em constrained}
$N=2$ superfields.  In the general case, we prove that the space
orthogonal to the kernel of the commutator of the complex
structures is integrable. This implies that a subsector of the
theory can be described as in the fully commuting case, but for
the remainder of the theory, as for the general (noncommuting)
case, no $N=2$ superspace  description is known. However, a
special case was described locally in \cite{BLR}.  Here we find an
application of the construction of \cite{BLR} by first showing that
the dual of the obvious complex structures on $SU(2)\times SU(2)$
gives rise to {\em non}commuting complex structures on $SU(2)\times
U(1)$, and then using these to construct a {\em new} $N=2$ superspace
action for the super WZW-model on $SU(2)\times U(1)$.

We begin by briefly reviewing some of the results of
\cite{GHR,BLR}. In $N=1$ superspace, additional supersymmetries
(beyond the one made manifest by virtue of being in superspace) are
generated by complex structures on the target space \cite{kahler}.
The condition that these give the usual supersymmetry algebra implies
that the complex structures are integrable, and the condition that
these are symmetries of the action implies that the metric is
hermitian with respect to the structures, and that they are
covariantly constant with respect to a connection with torsion that
is given in terms of the curl of a two form on the target space
\cite{GHR} (see also \cite{R} for a succinct review). The key
observation is that supersymmetry determines these properties, so if
we find a transformation of the target space that preserves the
supersymmetry, we will automatically find corresponding complex
structures. Duality is such a transformation, and thus we find dual
complex structures.

Extended supersymmetries are most naturally described in extended
superspace. However, it is not known how to do this in general. On
general dimensional grounds, an $N=2$ superspace lagrangian is
expected to be simply a potential function of some scalar
superfields. To get any dynamics at all, at least some of these
superfields must be constrained.  For example, if the target space
is K\"ahler, then the lagrangian is the K\"ahler potential, and the
superfields are chiral \cite{kahler}. In general, when the left and
the right complex structures ($J_\pm$ below) commute, then the
$N=2$ superspace lagrangian is known and is described in
terms of chiral and twisted chiral superfields \cite{GHR}.  However,
generically the left and right complex structures do not commute.
After considering some general aspects, we find an $N=2$ superspace
lagrangian for an example with {\em non}commuting complex
structures.

\section{Complex Structures and Duality}

Given a supersymmetric nonlinear $\sigma$-model lagrangian
\be
S= -\frac1{2\pi}\int D^{2}((g_{\mu \nu}+b_{\mu \nu})D_{+}X^{\mu}
 D_{-}X^{\nu})
\label{so}
\ee
with an isometry generated by a vector field $Y$ that obeys
$L_{Y}(db)=0$, one can choose (local) coordinates such that
$Y=\partial /\partial X^{0}$ and  the metric $g$ and torsion
potential $b$ are  independent of $X^{0}$. Then one can get a new
target manifold with new metric $\tilde{g}$ and torsion potential
$\tilde{b}$ by dualizing \cite{TB,FT} as follows:  Gauging the
symmetry with gauge fields $V_{\pm}$ (replace $D_{\pm}X^{0}$ with
$D_{\pm}X^{0}+V_{\pm}$ and choosing $X^{0}=0$ gauge), we get the first
order lagrangian
\ba
S =-\frac1{2\pi}\int D^{2}\left[ e_{00}V_{+}V_{-}+e_{i0}D_{+}X^{i}V_{-}
        +e_{0i}V_{+}D_{-}X^{i}
        +e_{ij}D_{+}X^{i}D_{-}X^{j}   \right. &{}& \nonumber  \\
\left.+\pht (D_{+}V_{-}+D_{-}V_{+}) \right] &{}& \ .
\label{si}
\ea
Here $\{\mu ,\nu\} =0,1,..,2n-1,\;\{i,j\}=1,..,2n-1,$ $e_{\mu\nu}
=g_{\mu\nu}+b_{\mu\nu}$, and $\pht $ is the lagrange multiplier whose
variation imposes $V_{\pm}=D_{\pm}X^{0}$ and gives back the original
action (\ref{so}). Extended $(2,2)$ supersymmetry of the action
(\ref{so}) implies that there is an invariance $\de_\e X^\mu= \e
J^\mu_{+\nu}D_+X^\nu $ and $\de_\eta X^\mu= \eta
J^\mu_{-\nu}D_-X^\nu$ where $J_\pm$ are integrable complex
structures that are covariantly constant with respect to a
connection with torsion $T_{\mu\nu\rho} =
\pm\frac12(\pa_\mu b_{\nu\rho}+\pa_\nu b_{\rho\mu}+\pa_\rho
b_{\mu\nu})$ \cite{GHR,HKLR,R}. The first order action (\ref{si}) is
invariant under
\ba
&\dep\pht=e_{\mu 0}\e
(J^{\mu}_{+0}V_{+}+J^{\mu}_{+j}D_{+}X^{j})\ , \ \ \
\dep X^{i}=\e(J^{i}_{+0}V_{+}+J^{i}_{+j}D_{+}X^{j})\ ,\ &\crr
&\dep V_+=D_+\dep X^0 \equiv
D_+[\e(J^0_{+0}V_{+}+J^0_{+i}D_{+}X^i)]\ , &
\ea
and
\ba
&\deta\pht = -e_{0\mu}\eta
(J^{\mu}_{-0}V_{-}+J^{\mu}_{-j}D_{-}X^{j})\ ,\ \ \
\deta X^{i}= \eta (J^{i}_{-0}V_{-}+J^{i}_{-j}D_{-}X^{j})\ ,\ &\crr
&\deta V_-=D_-\deta X^0
\equiv D_-[\eta(J^0_{-0}V_-+J^0_{-i}D_-X^i)]\ . &
\ea

To find the dual model, we eliminate $V_{\pm}$ by their equations of
motion
\be
V_{+}= e_{00}^{-1}(D_{+}\pht -e_{i0}D_{+}X^{i}) \ ,\ \ \
V_{-}=-e_{00}^{-1}(D_{-}\pht +e_{0i}D_{-}X^{i})
\label{vpm}
\ee
so that
\ba
\dep\pht &=&\e \left[ e_{\mu 0}J^{\mu}_{+0}e^{-1}_{00}D_{+}\pht
+(-e_{\mu 0}J^{\mu}_{+0}e^{-1}_{00}e_{i0}+e_{\mu
0}J^{\mu}_{+i}) D_{+}X^{i}\right]\ , \label{dep} \crr
\dep X^{j}&=&\e \left[ J^{j}_{+0}e^{-1}_{00}D_{+}\pht
+(J^{j}_{+i}-J^{j}_{+0}e^{-1}_{00}e_{i0})D_{+}X^{i} \right]\ ,\crr
\deta\pht &=&\eta \left[ e_{0\mu}J^{\mu}_{-0}e^{-1}_{00}D_{-}\pht+
(e_{0\mu}J^{\mu}_{-0}e^{-1}_{00}e_{0i}-e_{0\mu}J^{\mu}_{-i})D_{-}X^{i}
\right]\ ,\crr
\deta X^{j}&=&-\eta \left[J^{j}_{-0}e^{-1}_{00}D_{-}\pht
 -(J^{j}_{-i}-J^{j}_{-0}e^{-1}_{00}e_{0i})D_{-}X^{i}\right]\ .
\label{dex}
\ea
{}From the transformations (\ref{dep}-\ref{dex}), we can read off the
dual complex structures $K_{+}$ and $K_{-}$:
\be
K^{\mu}_{+\nu}= \left( \begin{array}{cc} e_{\mu
0}J^{\mu}_{+0}e^{-1}_{00} & -e_{\mu 0}J^{\mu}_{+0}e^{-1}_{00}e_{i0}
+e_{\mu 0}J^{\mu}_{+i}  \\ & \\ J^{j}_{+0}e^{-1}_{00}&
J^{j}_{+i}-J^{j}_{+0}e^{-1}_{00}e_{i0}
\end{array}  \right),
\ee
and
\be
K^{\mu}_{-\nu}= \left( \begin{array}{cc}
e_{0\mu}J^{\mu}_{-0}e^{-1}_{00} &
e_{0\mu}J^{\mu}_{-0}e^{-1}_{00}e_{0i} -e_{0\mu}J^{\mu}_{-i}  \\ & \\
-J^{j}_{-0}e^{-1}_{00}&  J^{j}_{-i}-J^{j}_{-0}e^{-1}_{00}e_{0i}
\end{array} \right)
\ee \\
with respect to $\{d\pht,\;dX^{1}\;..\;dX^{2n-1}\}$.  \\

Substituting (\ref{vpm}) into (\ref{si}), we get the new action [2]
\ba
\tilde S=-\frac1{2\pi}\int D^2 \left[\frac1{e_{00}}(D_{+}\pht D_{-}\pht
+e_{0j}D_{+}\pht D_{-}X^{j} -e_{i0}D_{+}X^{i}D_{-}\pht) \right.
&{}&\nonumber\\
\left.+(e_{ij}-{{e_{i0}e_{0j}}\over
{e_{00}}}) D_{+}X^{i}D_{-}X^{j}\right]&{}&
\label{sdual}
\ea
and read off the dual metric \~{g} and torsion potential \~{b}:
\be
\tilde{g}_{\mu \nu}=
\left( \begin{array}{cc}
g^{-1}_{00}&g^{-1}_{00}b_{0i}\\ & \\
g^{-1}_{00}b_{0i} & g_{ij}-g^{-1}_{00}(g_{i0}g_{j0}+ b_{i0}b_{0j})
\end{array}  \right),
\ee
and
\be
\tilde{b}_{\mu \nu}=
\left( \begin{array}{cc}
0 & g^{-1}_{00}g_{0i}\\ & \\
 -g^{-1}_{00}g_{0i} & b_{ij}+g^{-1}_{00}(g_{i0}b_{j0}-g_{j0}b_{i0})
\end{array}  \right)
\ee
with respect to the same basis as above. \\

We end this section with some observations. If $J_{+}=J_{-}$
(the K\"{a}hler case), then generally $K_{+}\neq K_{-}$, but
$[K_{+},K_{-}]=0$. Further, if we begin with $[J_{+},J_{-}]=0$ (and
$J_{+}\neq J_{-}$), then  in general $[K_{+},K_{-}]\neq 0$. In the
most general case, when $[J_{+},J_{-}]\neq 0$, the conditions for
$[K_+,K_-]=0$ are:
\be
\{J_+,J_-\}^i{}_0=[J_+,J_-]^\mu{}_0g_{\mu0}=
\{J_+,J_-\}^\mu{}_ig_{\mu 0}-\{J_+,J_-\}^\mu{}_0g_{\mu i}=0\ ,
\ee
\be
2(J_+^i{}_0J_{-0j}-J_-^i{}_0J_{+0j})=
g_{00}[J_+,J_-]^i{}_j-[J_+,J_-]^i{}_0g_{0j} \ ,
\ee \\
where $J_{0j}\equiv J^\mu{}_jg_{\mu 0}$.
\\

\section{Integrability}

As we have seen, the right and left complex structures $J_\pm$ on
the target manifold do not commute for generic $\sigma$-models with
torsion. A simple algebraic argument shows that the kernel of the
commutator $[\jp,\jm]$ is the direct sum of the kernels of the sum
and difference of the two complex structures:
\be
ker [\jp,\jm] = ker (\jp+\jm) \oplus ker (\jp-\jm)\ .
\ee
The whole manifold cannot be described in terms of chiral and
twisted chiral superfields. However, if there are one-forms
annihilated by $[\jp,\jm]$, and  if these one-forms form an
integrable Pfaff system (equivalently if $(ker[\jp,\jm])^\bot$ is
integrable), they can be integrated to coordinates on the
manifold. Here we show that the set of one-forms in the
kernel of $[\jp,\jm]$ is {\em always} integrable (we have also
shown that the one-forms in each $ker(\jp\pm\jm)$ are separately
integrable). These preferred coordinates are described by chiral
or twisted chiral superfields, and are perpendicular to a
foliation of the manifold by submanifolds on which the commutator
has no kernel.\footnote{In a previous version of this paper, we
concluded that the kernel was {\em not} integrable.  This was
because we looked at the integrability of of the {\em vector
fields} annihilated by $[\jp,\jm]$; this kernel is indeed not
integrable, which implies that there does not exist a foliation by
submanifolds on which $[\jp,\jm]=0$.  However, such a foliation is
not needed to guarantee the existence of chiral and twisted chiral
superfields.}

To prove that the forms in the kernel of the commutator
$[J_+,J_-]$ are integrable, it is convenient to choose
complex coordinates in which $J_+$ is diagonal and constant. The
vector index $\mu$ splits into holomorphic and anti-holomorphic
indexes $\mu \rightarrow m, \bar m$. In these coordinates, $J_+$ is
given by
\be
J_{+\bar n}^{m} \; = \; J_{+n}^{\bar m} \; =
\; 0\ , \ \ \ J_{+n}^{m} \; = i\delta^{m}_n \ ,
\ \ \ J_{+\bar n}^{\bar m} \; = \; -i\delta^{\bar m}_{\bar n}\ ,
\ee
and the components of the commutator are
\ba
&\left[ \jp , \jm \right]^{m}_{\;\; n} =
\left[ \jp , \jm \right]^{\bar m}_{\;\;\bar n} = 0 \ , &\label{poc}
\crr &\left[ \jp , \jm \right]^{m}_{\;\;\bar n} \,
= \, 2 i \; J_{-\bar n}^m\ \ ,\ \ \ \left[ \jp , \jm \right]^{\bar
m}_{\;\; n}  = - 2 i \, J_{-n}^{\bar m}\ . &
\ea
One-forms $a_{\mu} \; dx^{\mu}$ in the kernel of the commutator
satisfy
\be
a_\mu \left[ \jp , \jm \right]^\mu_{\;\;\nu} dx^\nu = 0\ ,
\label{kor}
\ee
which in the holomorphic coordinate system is equivalent to
\be
a_m J_{-\bar n}^{m} = a_{\bar m} J_{-n}^{\bar m} = 0\ .
\label{ker}
\ee

The underlying supersymmetric $\sigma$-model induces several
restrictions on the complex structures; as we use a
holomorphic coordinate system, only equations involving
$\jm$ are relevant. $J_-$ is covariantly constant
\be
D_{\mu} J_{-\nu}^\rho \equiv \partial_{\mu} J_{-\nu}^{\rho}  +
\Gamma^{(-)\rho}_{\;\;\lambda \mu} J_{-\nu}^{\lambda} -
\Gamma^{(-)\lambda}_{\;\; \nu \mu} J_{-\lambda}^{\rho} = 0
\label{jcovcons}
\ee
with respect to a connection with torsion
\be
\Gamma^{(-)\lambda}_{\;\; \mu \nu} = \{^{\lambda}_{\mu \nu}\} -
\frac12 T^{\lambda}_{\; \mu \nu}\ ,
\ee
where $\{^{\lambda}_{\mu \nu}\}$ is the Riemannian connection of the
metric on the target manifold, and $T^{\lambda}_{\; \mu \nu}$ is the
torsion tensor.  An algebraic identity relates the torsion and the
complex structure \cite{GHR},
\be
T^{\lambda}_{\; \mu \nu} = J_{-\mu}^{\rho} J_{-\nu}^{ \sigma}
T^{\lambda}_{\; \rho \sigma} + J_{-[ \nu |}^{\rho}
J_{-\sigma}^{\lambda} T^{\sigma}_{\; \rho |\mu ]}\ .
\label{alg}
\ee
Finally, because the metric is hermitian with respect to both complex
structures, in the holomorphic coordinate system for $\jp$ some of
the components of the connection $\Gamma^{(-)}$ and the torsion $T$
vanish:
\ba
&\Gamma^{(-)i}_{\;\; \bar j \bar k} =\Gamma^{(-)\bar i}_{\;\; j k} =
\Gamma^{(-)i}_{\;\; j \bar k} = \Gamma^{(-)\bar i}_{\;\; \bar j k} =
0\  ,&\label{gam} \crr
&T^{i}_{\; \bar j \bar k} = T^{\bar i}_{\; j k} = 0 \ .&
\label{tor}
\ea

Let $\omega^p = a^p_{\mu} dx^{\mu}, \; p=1,...,N$ span the
set one-forms in the kernel of the commutator
of the two complex structures. These one-forms can be
integrated simultaneously to $N$ coordinates on the manifold
$\cal{M}$ if and only if the Frobenius integrability condition is
satisfied, that is, if the algebra of the $dim({\cal M})-N$ vector
fields $X_q = X^{\nu}_q\frac{\partial}{\partial x^{\nu}}$
orthogonal to the forms $\omega^{p}$ closes. Explicitly, for
all vector fields $X$ satisfying $\omega^p(X)\equiv
a^p_{\mu}X^\mu=0$, we must have $[X_q,X_r]=C_{qr}{}^sX_s$.

Here the orthogonal vector fields $X$ are (see \ref{kor}):
\be
X_{\nu} \; = \; \left[ \jp , \jm \right]^{\mu}_{\;\;\nu} \;
\frac{\partial}{\partial  x^{\mu}}\ .
\ee
Note that the index $\nu$ on $X_\nu$ runs from $1,...,dim({\cal
M})$, but only $dim({\cal M})-N$ vector fields $X_{\nu}$ are linearly
independent. We have integrability if all the $\omega^p$ are
annihilated by $[X_\mu,X_\nu]$, {\it i.e.\/}, if
\be
a^p_{\mu} \; \left[ \jp , \jm \right]^{\mu}_{\;\; \nu}\ \
\Rightarrow\  \ a^p_{\mu} \;
\left[ \jp , \jm \right]^{\nu}_{\;\; [\kappa}
\; \left[\jp , \jm \right]^{\mu}_{\;\; \lambda ], \nu} \; = \; 0\ .
\label{int}
\ee

Consider first $\kappa, \lambda=k,l$ both
holomorphic. Then (\ref{poc}) restricts the summation over $\mu$ to
the antiholomorphic part: $\mu \rightarrow
\bar m$.  The integrability condition (\ref{int}) becomes
\be
a_{\bar m} \; J_{-[ k}^{\nu} \; J_{- l], \nu}^{\bar m} \; = \; 0\ .
\ee
We eliminate the derivatives on $\jm$ because it is covariantly
constant (\ref{jcovcons})
\be
a_{\bar m} J_{-[ k}^{\nu} \Gamma^{(-)\rho}_{\;\; l ] \nu}
J_{-\rho}^{\bar m} - a_{\bar m} J_{-[ k}^{\nu}
J_{-l ]}^{\rho} \Gamma^{(-)\bar m}_{\;\; \rho \nu} = 0\ .
\ee
In the first term, (\ref{ker}) implies that the summation over $\rho$
is only over the antiholomorphic part of that index. Because of this
and (\ref{gam}) we can antisymmetrize the $l$ and $\nu$ subscripts
of $\Gamma^{(-)}$ in the first term.  This picks the antisymmetric
part of the connection and we obtain:
\be
a_{\bar m} J_{-[k}^{\nu} T^{\rho}_{\; l] \nu} J_{-\rho}^{\bar m}
- a_{\bar m} J_{-k}^{\nu} J_{-l }^{\rho} T^{\bar m}_{\; \rho\nu}
= 0\ .
\ee
But using the identity (\ref{alg}) this is equivalent to
\be
a_{\bar m} T^{\bar m}_{\; k l } = 0\ ,
\ee
which is satisfied because of (\ref{tor}). The proof for $\lambda,
\kappa$ in (\ref{int}) both antiholomorphic is equivalent, and
we are left to consider, {\it e.g.\/}, $\kappa = \bar k$
antiholomorphic and $\lambda = l $ holomorphic. The integrability
condition (\ref{int}) becomes
\be
a_{\bar m} J_{-\bar k}^n J_{-nl,n}^{\bar m} -
a_m J_{-l}^{\bar n} J_{-\bar k,\bar n}^m = 0 \ .
\label{zep}
\ee
Each term in (\ref{zep}) vanishes by itself. The first one is
\be
a_{\bar m} J_{-\bar k}^{n} \left(\Gamma^{(-)\rho}_{\;\; l n}
J_{-\rho}^{\bar m} - \Gamma^{(-)\bar m}_{\;\; \rho n} J_{-l}^\rho
\right) = 0
\label{pop}
\ee
The summation over $\rho$ in the first term of (\ref{pop}) is
restricted to the antiholomorphic part because of (\ref{ker}) and
then the whole expression (\ref{pop}) vanishes because the components
of the connection entering it are zero (\ref{gam}). The
second term in (\ref{zep}) vanishes similarly.

Thus we have proven that the set of one-forms annihilated by
the commutator of the two complex structures can always be integrated
to coordinates $x^\mu$ on the manifold. They correspond to chiral and
twisted chiral superfields. The levels $x^\mu = constant$ of these
distinguished coordinates folliate the target manifold with
submanifolds on which the commutator of the two complex structures is
nondegenerate.

Actually, to conclude that the coordinates correspond to chiral and
twisted chiral multiplets, we need the integrability not just of
$(ker[\jp,\jm])^\bot$, but of $(ker(\jp\pm\jm))^\bot$ each
separately; this can be proven using the same techniques as above.
Then chiral superfields correspond to $ker(\jp-\jm)$ and twisted
chiral superfields to $ker(\jp+\jm)$.

\section{An example}
The $SU(2)\times SU(2)$ WZW model has the action $S=S_{1}+S_{2}$,
where
\be
S_{1}(h)=\frac1{2\pi}\int _{\pa M}tr(h^{-1}\pa h h^{-1}\bar{\pa}h)
+\frac1{2\pi}\int _{M}tr(h^{-1}dh\wedge h^{-1}dh\wedge h^{-1}dh)\ .
\ee
and similarly for $S_{2}$. Writing
$h=\exp^{\frac{i}{2}\thl\sigma_{3}}
\exp ^{\frac{i}{2}\phi \sigma _{2}}\exp ^{\frac{i}{2}\thr
\sigma_{3}}$where $h\in SU(2)$ and $\sigma _{1},\sigma _{2},
\sigma_{3}$ are Pauli matrices, we have
\be
S_{1}(h)=\frac1{2\pi}\int _{\pa M}(\pa\phi \bar{\pa}\phi
+\pa\thl\bar{\pa}\thl+\pa\thr\bar{\pa}\thr
+2\cos \phi \pa\thl\bar{\pa}\thr) \ .
\ee
The metric and the torsion potential can be read off from $S(h)$. We
construct complex structures on $SU(2)\times SU(2)$ starting
from the action on the Lie algebra $su(2)\oplus su(2)$; for example,
we may choose
\ba
&I_{+}:&\ E^{1,2}_{\pm}\rightarrow \pm iE^{1,2}_{\pm}\ ,\ \ E^{1}_{3}
\rightarrow E^{2}_{3}\ ,\ \ E^{2}_{3}\rightarrow -E^{1}_{3}\ ,
\\&{}&\nonumber\\
&I_{-}:&\ E^{1,2}_{\pm}\rightarrow \pm iE^{1,2}_{\pm}\ ,\ \ E^{1}_{3}
\rightarrow -E^{2}_{3}\ ,\ \ E^{2}_{3}\rightarrow E^{1}_{3}\ ,
\ea
where $I_{\pm}$ are expressed in a basis of left(right)-invariant
frames,  $e^{a}_{(L)},e^{a}_{(R)}$ defined by
$g^{-1}dg=e^{a}_{(L)}T_{a}$, $dgg^{-1}=e^{a}_{(R)}T_{a}$,
respectively. Here $T_{1},T_{2},T_{3}$ are $i/2$ times Pauli
matrices and $E^{1,2}_\pm =\frac{1}{\sqrt{2}}(T_{1}\pm iT_{2})\ ,\ \
E^{1,2}_{3}=T_{3}$. Thus we find the {\em non}commuting complex
structures
\be
J^{\mu}_{+\nu}=e^{\mu}_{(L)a}I^{a}_{+b}e^{b}_{(L)\nu}\ ,\;\;\;
J^{\mu}_{-\nu}=e^{\mu}_{(R)a}I^{a}_{-b}e^{b}_{(R)\nu}\ .
\label{jpm}
\ee

The commutator of the complex structures $J_\pm$ has a two dimensional
kernel.  We now find the corresponding complex coordinate explicitly.
In a basis of one forms arranged as $(d\theta^1_{L},d\theta^1_{R},
d\phi^1,d\theta^2_{L},d\theta^2_{R},d\phi^2)$, the two complex
structures are given by the following matrices:
\be
J_{+}=\left[ \begin{array}{cccccc}
       0 & 0 & -1/{\sin{\phi^1}} & 0 & 0 & 0 \\
       0 & 0 & \cot\phi^1 & \cos\phi^2 & 1 & 0 \\
       \sin\phi^1 & 0 & 0 & 0 & 0 & 0 \\
       0 & 0 & 0 & 0 & 0 & -1/{\sin\phi^2} \\
       -\cos\phi^1 & -1 & 0 & 0 & 0 & \cot\phi^2 \\
       0 & 0 & 0 & \sin\phi^2 & 0 & 0
             \end{array} \right]\ ,
\ee
\be
J_{-}=\left[ \begin{array}{cccccc}
       0 & 0 & -\cot\phi^1 & -1 & -\cos\phi^2 & 0 \\
       0 & 0 & 1/{\sin\phi^1} & 0 & 0 & 0 \\
       0 & -\sin\phi^1 & 0 & 0 & 0 & 0 \\
       1 & \cos\phi^1 & 0 & 0 & 0 & -\cot\phi^2 \\
       0 & 0 & 0 & 0 & 0 & 1/{\sin\phi^2} \\
       0 & 0 & 0 & 0 & -\sin\phi^2 & 0
             \end{array} \right]\ .
\ee
Explicit calculation shows that $J_{+}-J_{-}$ has no kernel but
$J_{+}+ J_{-}$ vanishes on a subspace of dimension two. The two
eigenforms are:
\ba
\frac1{\sin\phi^1}(\cos\phi^1 -1) d\phi^1 + d\theta^2_L +
d\theta^2_R \ ,\crr
\frac1{\sin\phi^2}(1-\cos\phi^2) d\phi^2 + d\theta^1_L +
d\theta^1_R\ .
\ea
These can be integrated to a complex coordinate $\eta$
\be
\eta = 2 \ln(\cos(\phi^1/2)) - \theta^2_L -\theta^2_R - 2i
\ln(\cos(\phi^2/2)) -i \theta^1_L -i \theta^1_R
\ee
that is holomorphic with respect to $\jp$ and
antiholomorphic with respect $\jm$:
\be
d\eta\; J_\pm =\pm id\eta\ .
\ee
In superspace, this coordinate is described by a twisted chiral
superfield.\\

We now perform a duality transformation with respect to a coordinate
$X^0=(\thr^1+\thl^1)/\sqrt2$ as described above in the general case,
and get a new metric, torsion, and complex structures
$K_{+}$ and $K_{-}$. The new space we get is $(SU(2)/U(1))\times
(SU(2)\times U(1))$. The two complex structures $K_{+}$ and
$K_{-}$ commute on an integrable 2 dimensional subspace of the tangent
bundle (the tangent space of the $SU(2)/U(1)$ factor), but do {\em not}
commute on $SU(2)\times U(1)$; the {\em anti}commutator of
$K_{+}$ and $K_{-}$  on $SU(2)\times U(1)$ gives $-2I\cos \phi$, where
$I$ is the identity. Explicitly, on $SU(2)/U(1)$, in a basis
$(d\phi^1,d\psi\equiv\frac{d\thr^1-d\thl^1}2-\frac{d\pht}{\sqrt2})$,
where
$\pht$ is the new dual coordinate (\ref{sdual}), we have:
\be
K_{+}=K_{-}=\left[ \begin{array}{cc}
       0 & -\tan\frac{\phi^1}2\\ & \\
       \cot\frac{\phi^1}2 & 0 \end{array}\right]\ .
\ee
On $U(1)\times SU(2)$, in a basis $(d\chi,d\thl^2,d\thr^2,\phi^2)$,
where $\chi\equiv\frac{\thr^1-\thl^1}2+\frac\pht{\sqrt2}$, we have:
\be
K_{+}=\left[ \begin{array}{cccc}
       0 & \cos\phi^2 &1& 0 \\
       0 & 0 & 0 & -1/\sin\phi^2  \\
       -1 & 0 & 0 & \cot\phi^2 \\
       0 & \sin\phi^2 & 0 & 0
       \end{array} \right]\ ,
\label{kplus}
\ee
\be
K_{-}=\left[ \begin{array}{cccc}
       0 & 1 & \cos\phi^2 & 0\\
       -1 & 0 & 0 & -\cot\phi^2 \\
       0 & 0 & 0 & 1/\sin\phi^2 \\
       0 & 0 & -\sin\phi^2 & 0
       \end{array} \right]\ .
\label{kminus}
\ee
On $SU(2)/U(1)$ there is no torsion and the metric is
\be
\tilde{g}=\left[ \begin{array}{cc}1 & 0 \\ 0 & \tan ^{2} \frac{\phi^1}2
\end{array} \right]\ ,
\ee
and on $U(1)\times SU(2)$ the metric and torsion potential are
\be
\tilde{g}=\left[ \begin{array}{cccc}
       1 & 0& 0 & 0 \\
       0 & 1 & \cos\phi^2 & 0 \\
       0 & \cos\phi^2 & 1 & 0 \\
       0 & 0 & 0 & 1
         \end{array} \right]\ ,
\ee
\be
\tilde{b}=\left[ \begin{array}{cccc}
       0 & 0 & 0 & 0 \\
       0 & 0 & \cos\phi^2 & 0 \\
       0 & -\cos\phi^2 & 0 & 0 \\
       0 & 0 & 0 & 0
         \end{array} \right]\ .
\ee
\\

The essential point is that though $SU(2)\times U(1)$ admits
commuting left and right complex structures, duality has given us
some others. This can be most easily understood if we
paramatrize the group $SU(2)\times U(1)$ as (see, for example,
\cite{RSS}):
\be
 g=\frac{e^{i\theta }}{\sqrt {\Phi \bar{\Phi}+\Lambda \bar{\Lambda }}}
\left( \begin{array}{cc}
    \Lambda & \bar{\Phi} \\
    -\Phi  & \bar{\Lambda}
    \end{array} \right) , \;\;\;\;\;
\theta =-\frac{1}{2}\ln (\Phi\bar{\Phi}+\Lambda\bar{\Lambda}).
\ee
Letting the imaginary quaternions $\left( \begin{array}{cc}
0& i\\ i & 0 \end{array} \right),\left( \begin{array}{cc}
0& -1\\ 1 & 0 \end{array} \right),\left( \begin{array}{cc}
i&0\\ 0 &-i \end{array} \right) $ act on $\left( \begin{array}{cc}
d\Lambda  & d\bar\Phi \\ -d\Phi & d\bar{\Lambda} \end{array} \right)$
from left or right gives two commuting sets of complex structures
$J^{1}_{-},J^{2}_{-},J^{3}_{-}$ and
$J^{1}_{+}, J^{2}_{+},J^{3}_{+}$; in terms of these, $K_{+}$ and
$K_{-}$ (\ref{kplus},\ref{kminus}) are expressed as
\be
K_{+}=J^{3}_{+},\;\;\;\;\;\; K_{-}=\cos \phi^2 J^{3}_{+}+\sin \phi^2
\cos\thr^2 J^{1}_{+}-\sin \phi^2 \sin \thr^2 J^{2}_{+}
\ee
and it is clear their anticommutator is $-2\cos \phi^2$. We can
interpret this as follows: by construction, $[J_{+},J_{-}]=0$.
However, $SU(2)\times U(1)$ has an orientation reversing outer
automorphism $\Omega$ that acts by taking the $U(1)$ generator
$Q\rightarrow -Q$. This preserves $g$ and $b$, but not $J$; so
$K_{-}=\Omega (J^{3}_{+}).$

We summarize the situation as follows: On $SU(2)\times
SU(2)$, there is no choice of $J_+$ and $J_-$ such that
$[J_+,J_-]=0$. The dual of $SU(2)\times SU(2)$
is $SU(2)\times U(1) \times SU(2)/U(1)$; this {\em does} admit
$J_\pm$ that commute, but the complex structures that we find by
duality ($K_\pm$ above) do {\em not} commute.

\section{A new $N=2$ description of $SU(2)\times U(1)$}
The essential ingredient for $N=2$ superspace in two dimensions
(with Lorentzian signature) is a pair of complex
spinor derivatives $D_\pm$ that obey the algebra (see, for example,
\cite{GHR,R}):
\be
D_\pm^2=\bar D_\pm^2 = \{ D_\pm,\bar D_\mp\}=0\ ,\qquad \{
D_\pm,\bar D_\pm\} = \pm i \partial_\pm\ ,
\ee
where $\partial_\pm$ are the usual two-dimensional
derivatives along the two lightlike directions. An $N=2$ superspace
action is written as
\be
S=\frac1{2\pi}\int D^2\bar D^2 K(\Phi^i)\ ,
\ee
where $K$ is some potential function of various superfields
$\Phi^i$. As discussed in the introduction, a
$(2,2)$ superfield description of arbitrary $N=2$ supersymmetric
sigma-models (in particular,
WZW-models) is not known {\em unless} $J_\pm$ commute. In that
case, the model can be described in terms of chiral superfields
$\Phi$ ($\bar D_\pm \Phi =0$) and twisted chiral superfields
$\Lambda$ ($\bar D_+\Lambda =D_-\Lambda=0$) [9]. The simplest
example is $SU(2)\times U(1)$, where we parametrize the group
element as \cite{RSS}
\be
 g=\frac{e^{i\theta }}{\sqrt {\Phi \bar{\Phi}+\Lambda \bar{\Lambda }}}
\left( \begin{array}{cc}
    \Lambda & \bar{\Phi} \\
    -\Phi  & \bar{\Lambda}
    \end{array} \right),\qquad
 \theta =-\frac{1}{2}\ln (\Phi\bar{\Phi}+\Lambda\bar{\Lambda})\ ,
\ee
and the superspace Lagrangian is
\be
K=-\int^{\frac{\Lambda \bar{\Lambda}}{\Phi \bar{\Phi}}}
  \frac{d\zeta}{\zeta}\ln (1+\zeta )+\frac{1}{2}(\ln
(\Phi \bar{\Phi}))^{2}\ .
\label{eq:suplag}
\ee
By construction, the manifest complex structures are the commuting
ones. We now give an $N=2$ superspace description with respect
to the noncommuting complex structures $K_\pm$ we found above.
\\

To do this,
we use a different superspace description in terms of semichiral and
anti-semichiral superfields
$\Phi_{1},\bar{\Phi}_{1},\Phi_{2},\bar{\Phi}_{2}$ (which obey $\bar
D_+\Phi_1=\bar D_- \Phi_2=D_+\bar\Phi_1=D_-\bar\Phi_2=0$)
\cite{BLR}. We find the action by a new kind of duality
transformation. Then by construction we have the same metric and
torsion, but different complex structures.
More explicitly, we add lagrange multiplier terms to the lagrangian
(\ref{eq:suplag})
to get the first order lagrangian,\footnote{This duality
superficially resembles a well-known and trivial duality between
chiral (twisted chiral) superfields and linear (twisted linear)
superfields.}
\begin{eqnarray}
K & = & -\int^{x}\frac{d\zeta}{\zeta}\ln (1+\zeta )
								+\frac{1}{2}(\ln y)^{2}
    +(\Phi_{1}+\Phi_{2})e^{i\theta}y
    +(\bar{\Phi}_{1}+\bar{\Phi}_{2})e^{-i\theta}y \nonumber \\
 &  &\qquad\qquad +(\Phi_{1}+\bar{\Phi}_{2})xye^{i\phi}
    +(\bar{\Phi}_{1}+\Phi_{2})xye^{-i\phi}
\label{kfirst}
\end{eqnarray}
where
\be
x=\frac{\Lambda \bar{\Lambda }}{\Phi \bar{\Phi }},\;\;\;\;\;
y=\Phi \bar{\Phi },\;\;\;\;\;
e^{i\theta}=\frac{\Phi }{\bar{\Phi }},\;\;\;\;\;
e^{i\phi}=\frac{\Lambda }{\bar{\Lambda }}\ ,
\ee
(here $\Phi ,\bar{\Phi},\Lambda ,\bar{\Lambda}$, and hence
$x,y,\theta,\phi$, are now unconstrained superfields).

As $\Phi_{1}$ is semichiral ($\bar{D}_{+}\Phi_{1}=0$),
$\Phi_{1}=\bar{D}_{+}\psi_{-}$ for some unconstrained $\psi_{-}$, and
varying $\psi_{-}$ we get
\be
\bar{D}_{+}(\Lambda +\Phi )=0\ .
\ee
Similarly, we have
\be
\bar{D}_{-}(\Phi +\bar{\Lambda })=0\ ,
\ee
\be
D_{+}(\bar{\Lambda }+\bar{\Phi })=0\ ,
\ee
\be
D_{-}(\bar{\Phi}+\Lambda )=0\ .
\ee

{}From the above, we get
\be
\bar{D}_{+}D_{-}(\Phi-\bar{\Phi})=0\ ,
\ee
\be
D_{+}\bar{D}_{-}(\Phi -\bar{\Phi})=0\ ,
\ee
which show that $\Phi$ is chiral superfield. Similarly $\Lambda $ is
shown to be twisted chiral superfield, and we recover the superspace
Lagrangian (\ref{eq:suplag}). Instead, if we vary
$\theta,\phi,x,y$, we find the relations
\be
e^{i\theta}=(\frac{\bar{\Phi}_{1}+\bar{\Phi}_{2}}{\Phi_{1}+\Phi_{2}})
 ^{\frac{1}{2}}
\ee
\be
e^{i\phi}=(\frac{\bar{\Phi}_{1}+\Phi_{2}}{\Phi_{1}+\bar{\Phi}_{2}})
 ^{\frac{1}{2}}
\ee
\be
-\frac{\ln (1+x)}{x}+2y|\Phi_{1}+\bar{\Phi}_{2}|=0
\ee
\be
\frac{1}{y}\ln y +2|\Phi_{1}+\Phi_{2}|+2x|\Phi_{1}+\bar{\Phi}_{2}|=0.
\ee
In this way we can express $\theta,\phi,x,y$ in terms of
$\Phi_{1},\bar{\Phi}_{1},\Phi_{2},\bar{\Phi}_{2}$ (at least
implicitly).

Substituting back into (\ref{kfirst}) gives a new potential
$K(\Phi_{1},\bar{\Phi}_{1},\Phi_{2},\bar{\Phi}_{2})$, and we can find
the corresponding complex structures; in particular the
anticommutator of these complex structures is proportional to the
identity \cite{BLR}. Our dual complex structures
$K_{+}$ and $K_{-}$ are both hermitian and covariantly constant and
their anticommutator has the same property as above.

$N=2$ superspace encodes complex structures: For example, chiral and
twisted chiral superfields give complex structures $J_{\pm}$ with
$[J_{+},J_{-}]=0$ \cite{GHR}. However, generic compact $WZW$ models
have $[J_{+},J_{-}]\neq 0$. The orientation reversed complex
structures $K_{\pm}$ on $SU(2)\times U(1)$ give an example since
$[K_{+},K_{-}]\neq 0$. We have found and presented an $N=2$
superspace description based on these complex structures, and hope
that this may give insight into the general case.

\begin{flushleft}
{\bf Acknowledgments}
\end{flushleft}

It is a pleasure to thank Ulf Lindstr\"om for reading the
manuscript and making many useful suggestions. We acknowledge partial
support from the NFS under Grant No.\ PHY 93 09888.\\


\begin{thebibliography}{6666}

\newcommand{\np}{Nucl.\ Phys.\ }
\newcommand{\pr}{Phys.\ Rev.\ }
\newcommand{\cmp}{Commun.\ Math.\ Phys.\ }
\newcommand{\pl}{Phys.\ Lett.\ }
\bibitem{FT}
E.S.\ Fradkin \& A.A.\ Tseytlin, Ann. Phys. {\bf 162} (1985) 31.
\bibitem{TB}
T.\ Buscher, \pl {\bf B159} (1985) 127,
\pl {\bf B194} (1987) 59,
\pl {\bf B201} (1988) 466.
\bibitem{HKLR}
N.\ J.\ Hitchin, A.\ Karlhede, U.\ Lindstr\"{o}m \&
M.\ Ro\v{c}ek, \cmp {\bf 108} (1987) 535.
\bibitem{RV}
M.\ Ro\v cek \& E.\ Verlinde, \np {\bf B373} (1992) 630.
\bibitem{duality}
A.\ Giveon, E.\ Rabinovici \& G.\ Veneziano, Nucl. Phys. {\bf B322}
(1989) 167;\\
A.\ Shapere \& F.\ Wilzcek, \np {\bf B320} (1989) 669;\\
A.\ Giveon, N.\ Malkin \& E.\ Rabinovici, \pl {\bf B220}
(1989) 551.
\bibitem{more}
S.\ Ferrara, D.\ L\"ust, A.\ Shapere \& S.\ Theisen, \pl
{\bf B233} (1989) 147;\\
J.\ Lauer, J.\ Maas \& H.P.\ Nilles, \pl
{\bf B226} (1989) 251; \np {\bf B351} (1991) 353;\\
W.\ Lerche, D.\ L\"ust \& N.P.\ Warner,  \pl
{\bf B231} (1989) 418;\\
S.\ Ferrara, D.\ L\"ust \& S.\ Theisen, \pl {\bf B233} (1989) 147;\\
A.\ Giveon, N.\ Malkin \& E.\ Rabinovici, \pl {\bf B238} (1990) 57;\\
A.\ Font, L.\ Ibanez, D.\ Lust \& F. Quevedo,
\pl {\bf B245} (1990) 401;\\
A.\ Giveon \& D.-J.\ Smit, \np {\bf B349} (1991) 168;\\
A.\ Giveon \& M.\ Porrati, \pl {\bf B246} (1990) 54; \np
{\bf B355} (1991) 422;\\
P.\ Candelas, X.C.\ de la Ossa, P.S.\ Green \& L.\ Parkes, \np {\bf
B359} (1991) 21;\\
T.\ Kugo \& B.\ Zwiebach, Prog.\ Theor.\ Phys.\ {\bf 87} (1992) 801.
\bibitem{GR}
A.\ Giveon \& M.\ Ro\v cek, \np {\bf B380} (1992) 128.
\bibitem{GHR}
S.\ J.\ Gates, C.\ M.\ Hull \& M.\ Ro\v cek,
\np {\bf B248} (1984) 157.
\bibitem{BLR}
T.\ Buscher, U.\ Lindstr\"om \& M.\ Ro\v cek.,
\pl {\bf B202} (1988) 94.
\bibitem{R}
M.\ Ro\v cek, {\it Modified Calabi-Yau Manifolds with Torsion}, in
`Essays on Mirror Manifolds', ed. S.\ T.\ Yau (International Press,
Hong Kong, 1992) 480.
\bibitem{kahler}
B.\ Zumino, \pl {\bf B87} (1979) 203; \\
L.\ Alverez-Gaum\' e \& D.\ Z.\ Freedman, \cmp {\bf 80} (1981) 443.
\bibitem{BBK}
B.\ B.\ Kim, SUNY at Stony Brook Ph.D.\ Thesis.
\bibitem{RSS}
M.\ Ro\v cek, K.\ J.\ Schoutens \& A.\ Sevrin.,
\pl {\bf B265} (1991) 303.
\end{thebibliography}
\end{document}